\documentclass{article}
\usepackage{spconf,amsmath,graphicx}
\usepackage{amssymb}
\usepackage{siunitx}
\usepackage{enumitem}
\usepackage{placeins}
\newcommand{\C}{\mathbb{C}}

\newcommand{\mbf}[1]{\mathbf{#1}}

\newcommand{\absv}[1]{\left\lvert{#1}\right\rvert}
\newcommand{\norm}[1]{\left\|{#1}\right\|}
\newcommand{\identity}{\mathbf{I}}
\renewcommand{\H}{\mathsf{H}}
\newcommand{\T}{\mathsf{T}}

\title{DIRECTIONAL SPARSE FILTERING USING WEIGHTED LEHMER MEAN\\FOR BLIND SEPARATION OF UNBALANCED SPEECH MIXTURES}
\name{
    Karn Watcharasupat, 
    Anh H. T. Nguyen, 
    Ching-Hui Ooi, and
    Andy W. H. Khong
\thanks{The first author was supported by the CN Yang Scholars Programme, Nanyang Technological University, Singapore. The remaining authors were supported within the Delta-NTU Corporate Lab for Cyber-Physical Systems with funding support from Delta Electronics Inc and the National Research Foundation (NRF) Singapore under the Corp Lab@University Scheme (Ref. SLE-RP5) at Nanyang Technological University, Singapore.}
\thanks{The source code for the proposed algorithm is available at \texttt{github.com/e13000/directional\_sparse\_filtering}.}
}
\address{School of Electrical \& Electronic Engineering\\
Nanyang Technological University, Singapore\\
Email: karn001@e.ntu.edu.sg, \{nguyenhta, chinghui.ooi, andykhong\}@ntu.edu.sg}
\begin{document}
\topmargin=0mm 
\sisetup{detect-all}
%
\maketitle
\begin{abstract}
    In blind source separation of speech signals, the inherent imbalance in the source spectrum poses a challenge for methods that rely on single-source dominance for the estimation of the mixing matrix. We propose an algorithm based on the directional sparse filtering (DSF) framework that utilizes the Lehmer mean with learnable weights to adaptively account for source imbalance. Performance evaluation in multiple real acoustic environments show improvements in source separation compared to the baseline methods.
\end{abstract}
\section{Introduction}
\label{sec:intro}

Unsupervised blind source separation (BSS) is the process of extracting source signals from its mixture with little to no prior information about the sources and without prior training using labelled data. In this paper, we focus on the problem of estimating the complex-valued mixing matrix from a multichannel observed mixture, particularly that of speech signals. We assume that the data, at each frequency bin, follow the noiseless linear mixing model
\begin{equation}
    \mbf{x}[k] = \mbf{A}\mbf{s}[k], \ k = 0, \dots, K-1,
\end{equation}
where $k$ is the time frame index, $K$ is the total number of frames, $\mbf{x}[k] \in \C^{M\times1}$ is the observed mixture signals acquired using $M$ microphones, $\mbf{s}[k] \in \C^{N \times 1}$ is the unknown source signals from $N$ speech sources, and $\mbf{A} \in \C^{M\times N}$ is the mixing matrix to be estimated. Similar to \cite{nguyen2017learning, nguyen2020directional}, we assume that the number of sources $N$ is known in this model.

The nature of the speech signal poses a unique challenge to the task of mixing matrix estimation. Since the natural speech volume differs across individuals, the power of each source signal also varies. Moreover, the relative energy of the source spectra also varies across frequencies. Existing directional sparse filtering (DSF) algorithms \cite{nguyen2017learning, nguyen2020directional}, however, implicitly assume that all sources have equal proportion of active time at each frequency bin of the mixture and, therefore, their estimation performance reduces in the presence of source spectra variation. 

Inspired by the weighted scheme in directional clustering algorithms based on mixture models \cite{vu2010blind, ito2016complex}, we propose an extension of the DSF algorithm that is robust to variations in the source spectra. To achieve such robustness, we employ the weighted Lehmer mean \cite{lehmer1971compounding} with learnable weights such that the variation in the source spectra are adaptively accounted for during the learning process. The proposed algorithm requires approximately the same computational complexity compared to the original unweighted DSF. In addition, the proposed algorithm does not require any constraints on the scale of the weights. As opposed to existing mixture models, the proposed method also does not assume the sources to be Gaussian distributed and, as a result, is more suitable for sparse sources such as speech signals that are super-Gaussian in nature.

\section{THE PROPOSED DSF-WLM ALGORITHM}
\label{sec:format}

\subsection{Directional Sparse Filtering with Lehmer Mean}
It is well known that speech signals are highly sparse in the time-frequency domain \cite{vincent2007complex}. Consequently, they have been observed to be approximately disjoint-orthogonal in the time-frequency domain \cite{bofill2001underdetermined, rickard2002approximate, simoncelli1999modeling}. Due to this property, one of the sources is often dominant in the mixture at each time-frequency bin. Directional clustering methods, such as $K$-hyperlines (KHL) \cite{o2004hard, he2009k}, directional mixture models \cite{vu2010blind, sawada2010underdetermined}, and directional sparse filtering \cite{nguyen2020directional}, exploit the sparseness constraint to estimate the mixing matrix $\mbf{A}$. 
We define $\mbf{H} = [\mbf{h}_1, \dots \mbf{h}_N]$ to be the estimate of $\mbf{A}$, such that $\mbf{h}_n$ denotes the $n$th column of $\mbf{H}$. KHL, in particular, estimates the mixing matrix by clustering the data based on the phase-invariant cosine distance
\begin{equation}
    \mathcal{D}_{n, k}
        = 1 - \dfrac{
                \absv{\mbf{h}_n^\H \mbf{x}[k]}^2
            }{
                \norm{\mbf{h}_n}^2 \norm{\mbf{x}[k]}^2
            },
    \label{eq:dist}
\end{equation}
where $(\cdot)^\H$ is the Hermitian transpose and $\norm{\cdot}$ is the $L_2$-norm. Directional sparse filtering is an extension of the KHL algorithm and seeks to minimize the cost function \cite{nguyen2020directional}
\begin{equation}
    \mathcal{J}^{\text{(PM)}}(\mbf{H})
    = \dfrac{1}{K}\sum_{k=0}^{K-1}\left(\frac{1}{N}\sum_{n=1}^{N}\mathcal{D}_{n,k}^p\right)^{1/p},
    \label{eq:pmcost}
\end{equation}
where $p<0$ is a chosen hyper-parameter and PM denotes for the power mean. As seen from \eqref{eq:pmcost}, the cost function is unweighted and hence, estimation performance may be reduced when the source activities are not uniformly distributed across the frames in a particular frequency bin.

To address the above issue, we propose the following cost function for  directional sparse filtering by weighted Lehmer mean (DSF-WLM)
\begin{equation}
    \mathcal{J}(\mbf{H},\mbf{w})
        =\dfrac{1}{K}\sum_{k=0}^{K-1}
            \dfrac{
                \sum_{n=1}^{N}\max(w_{n}+\alpha,\alpha)\mathcal{D}^{r}_{n,k}
            }{
                \sum_{n=1}^{N}\max(w_{n}+\alpha,\alpha)\mathcal{D}^{r-1}_{n,k}
            },
    \label{eq:lmcost}
\end{equation}
where $\mbf{w}=[w_{1},\dots,w_{N}]^{\T}\in\mathbb{R}^{N\times1}$ is the weight vector, $r$ is a hyper-parameter that controls the interpolation property, and $\alpha\ge0$ is a smoothing hyper-parameter. The hyper-parameter $\alpha$ specifies the minimum effective weights for each mixing filter to be estimated such that when $\alpha = 0$, no smoothing is performed. On the other hand, when $\alpha \gg \max_n w_n$, the effective weights are approximately equivalent, resulting the cost function being approximately unweighted.

To gain a better understanding of the proposed cost function, we first define the cosine-squared angular vector between the current estimated mixing matrix and the $k$th frame of the data by
\begin{equation}
    \mbf{u}_k = \begin{bmatrix}
        \dfrac{
                \absv{\mbf{h}_1^\H \mbf{x}[k]}^2
            }{
                \norm{\mbf{h}_1}^2 \norm{\mbf{x}[k]}^2
            },
        \dots,
        \dfrac{
                \absv{\mbf{h}_N^\H \mbf{x}[k]}^2
            }{
                \norm{\mbf{h}_N}^2 \norm{\mbf{x}[k]}^2
            }
        \end{bmatrix}^\T.
\end{equation} 
Note that $u_{n,k}$ (defined as the $n$th component of $\mbf{u}_k$) is related to the distance metric by $\mathcal{D}_{n,k} = 1- u_{n,k}$. In this paper, the hyper-parameter $r$ is restricted to the range $0 < r < 1$, where the Lehmer mean satisfies the three criteria for a sparse penalty function outlined in \cite{nguyen2020directional}:
\begin{enumerate}[leftmargin=*]
    \item The Lehmer mean decreases when the largest element of $\mbf{u}_k$ increases; this is an inherent soft-minimum property of the Lehmer mean when $r < 1$. When a component of $\mbf{u}_k$ approaches unity, this property allows DSF to behave in a manner similar to that of directional clustering.
    \item For $r < 1$, it can be shown that the Lehmer mean of $\mbf{1} - \mbf{u}_k$ is zero when the largest element of  $\mbf{u}_k$ is unity, regardless of other elements of the angular vector. This property ensures that single-source frames are treated equally. 
    \item For equal weights, the Lehmer mean of $\mbf{1} - \mathbf{u}_{k}$ is strictly Schur-concave with respect to $\mathbf{u}_{k}$ for $0 < r < 1$. This can be shown by noting the strict Schur-concavity of the Lehmer mean for $0 < r < 1$ \cite{fub2016schur}, and that the composition between a strictly Schur-concave function and a linear function is strictly Schur-concave.  Due to the sparsity enforcing property of Schur-concave functions \cite{kreutz2003dictionary}, the unwanted directional information from non-dominant source will be suppressed. We note that the proposed method considers a relaxation of this property since the weighted Lehmer mean may not be strictly Schur-concave.
\end{enumerate}

\subsection{Comparison to the weighted power mean}
We remark that a weighted extension of DSF by the weighted power mean (WPM) exists and is given by
\begin{equation}
    \mathcal{J}^{\text{(WPM)}}(\mbf{H}, \mbf{w})
    = \dfrac{1}{K}\sum_{k=0}^{K-1}\left(\dfrac{\sum_{n}\max(w_{n}+\alpha,\alpha)\mathcal{D}_{n,k}^p}{N\sum_{i=1}^N\max(w_{i}+\alpha,\alpha)}\right)^{\frac{1}{p}}.
    \label{eq:wpmcost}
\end{equation}
Here, the weights must be normalized by $\sum_{i}\max(w_{n}+\alpha,\alpha)$ so that the effective weights sum to unity. This normalization significantly complicates the partial gradient with respect to the weights. As opposed to the cost function given in \eqref{eq:wpmcost}, our proposed cost function in \eqref{eq:lmcost} offers a simpler gradient, thus reducing computational complexity. Moreover, the weighted Lehmer mean offers an additional benefit of weight self-normalization, i.e., the function is only affected by the relative difference between the weights but not the scale of the weights. This makes the weights of Lehmer mean easy to learn with minimal increase in computational demand compared to the unweighted version. 

\subsection{Optimization}
To facilitate estimation, we first perform whitening on $\mbf{x}$ such that it is zero-mean and uncorrelated with identity covariance, followed by $L_2$-norm normalization such that $\|\mbf{x}[k]\| = 1, \ \forall k$. By assuming that the source signals are also zero-mean and uncorrelated, we force the mixing matrix to be approximately semi-unitary, i.e., $\mbf{A}\mbf{A}^\H \approx \identity_M$. Since the weights of the Lehmer mean is self-normalizing, constraints on the scale of the weights are not required. As such, we arrive at the constrained optimization problem
\begin{equation}
    \min_{\mbf{H}, \mbf{w}}\mathcal{J}(\mbf{H}, \mbf{w}) \ \text{s.t.} \ \mbf{H}\mbf{H}^\H = \identity_M.
    \label{eq:optim}
\end{equation}
Let $\Re(\cdot)$ and $\Im(\cdot)$ denote the real and imaginary parts of a matrix, respectively. The gradient of the cost function with respect to the $n$th column of the mixing matrix $\mbf{h}_n$ is therefore given by
\begin{align}
    \nabla_{\mbf{h}_n^\ast}{\mathcal{J}}
        &= \dfrac{\mbf{g}_n}{\|\mbf{h}_n\|} - \dfrac{\mbf{h}_n}{\|\mbf{h}_n\|^3}\Re\left[\mbf{h}_n^\H \mbf{g}_n\right],
\end{align}
where 
\begin{align}
    \mbf{g}_n 
        &= \dfrac{2}{K}\sum_{k=0}^{K-1}\mathcal{B}_{n,k}\left(\dfrac{\mbf{x}^\H[k]\mbf{h}_n}{\|\mbf{h}_n\|}\right)\mbf{x}[k],\\
    \mathcal{B}_{n,k} 
        &= \dfrac{
            (r-1)\mathcal{D}^{r-2}_{n,k}\mathcal{L}_{k} - r\mathcal{D}^{r-1}_{n,k}
        }{
            \sum_{i=1}^{N}\max(w_{i}+\alpha,\alpha)\mathcal{D}^{r-1}_{i,k}
        },\\
    \mathcal{L}_{k} 
        &= \dfrac{\sum_{i=1}^{N}\max(w_{i}+\alpha,\alpha)\mathcal{D}^{r}_{i,k}}{\sum_{i=1}^{N}\max(w_{i}+\alpha,\alpha)\mathcal{D}^{r-1}_{i,k}}.
\end{align}
The gradient with respect to the weights is given by
\begin{align}
     \nabla_{w_n}{\mathcal{J}}
     &= 
     \begin{cases}
     \dfrac{1}{K}\displaystyle\sum_{k=0}^{K-1}\dfrac{
            \mathcal{D}^r_{n,k}
                    - \mathcal{D}^{r-1}_{n,k}\mathcal{L}_k
        }{
            \sum_{i}(w_i+\alpha)\mathcal{D}^{r-1}_{i,k}
        }, & \text{if $w_n > 0$,}\\
    0, & \text{otherwise.}
     \end{cases}
\end{align}

Satisfying the semi-unitary constraint is, unfortunately, not trivial. In order to simplify the problem, we substitute the semi-unitary constraint projection directly into the cost function \eqref{eq:lmcost}. Note that, in practice, it is also possible to omit the semi-unitary constraint with modest performance loss. We now arrive at the unconstrained optimization problem
\begin{equation}
    \min_{\widetilde{\mbf{H}}, \mbf{w}}\mathcal{J}\left((\widetilde{\mbf{H}}\widetilde{\mbf{H}}^\H)^{-1/2}\widetilde{\mbf{H}}, \mbf{w}\right),
    \label{eq:uoptim}
\end{equation}
where $\widetilde{\mbf{H}}$ is an auxiliary variable \cite{nguyen2020directional}. The optimization problem in \eqref{eq:uoptim} can now be easily solved with most gradient methods. Once the optimal auxiliary matrix $\widetilde{\mbf{H}}$ is found, we compute the estimated mixing matrix by normalizing $\widetilde{\mbf{H}}$ to satisfy the unitary constraint.

It can be seen that solving the optimization problem in \eqref{eq:uoptim} involves minimization of the real-valued cost function with respect to a complex-valued matrix $\widetilde{\mbf{H}}$ and a real-valued vector $\mbf{w}$. Similar to \cite{nguyen2020directional}, we minimize \eqref{eq:uoptim} using the limited-memory Broyden–Fletcher–Goldfarb–Shanno (L-BFGS) algorithm \cite{liu1989limited}. During optimization, the real and imaginary parts of $\widetilde{\mbf{H}}$ are treated as separate variables, that is, $\mathrm{vec}(\Re (\widetilde{\mbf{H}}))$, $\mathrm{vec}(\Im (\widetilde{\mbf{H}}))$, and $\mbf{w}$ are concatenated into a single parameter vector. The mixing matrix $\widetilde{\mbf{H}}$ is initialized using KHL and all weights are initialized to $w_n = {K + (N-1)\alpha}, \ \forall n$.

\begin{figure}[t]
    \centering
    \includegraphics[width=6cm]{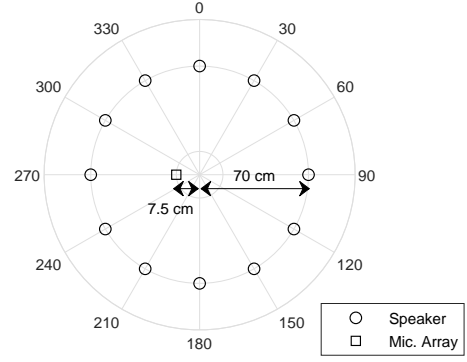}
    \caption{Experiment setup. The distances are measured relative to the center of the setup.}
    \label{fig:setup}
\end{figure}
\begin{figure}[t]
    \centering
    \includegraphics[width=8.5cm]{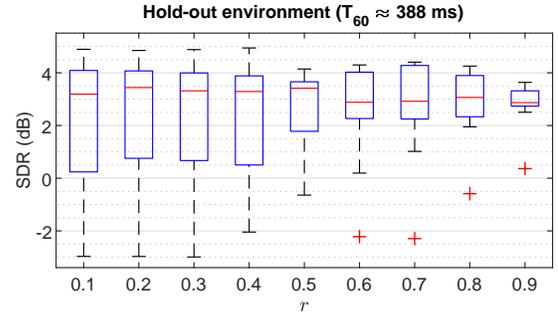}
    \caption{SDR performances of DSF-WLM with respect to $r$ with fixed $\alpha = 10$ in the hold-out environment.}
    \label{fig:grid}
\end{figure}

\section{SIMULATION RESULTS}
\label{sec:majhead}

\begin{figure*}[!th]
    \centering
    \includegraphics[width=0.95\textwidth]{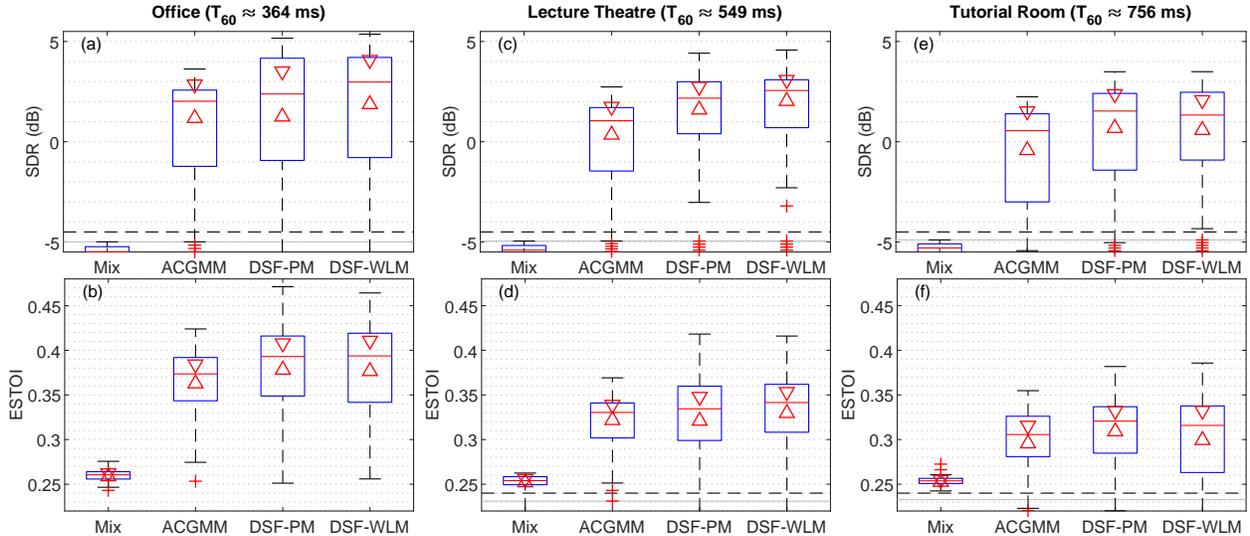}
    \caption{Performance of ACGMM, DSF-PM, and the proposed DSF-WLM algorithm. The triangular markers indicate the comparison intervals for the median performance. (a) and (b): SDR and ESTOI for office environment. (c) and (d): SDR and ESTOI for lecture theatre environment. (e) and (f): SDR and ESTOI for tutorial room environment.}
    \label{fig:results}
\end{figure*}

We compared the performance of the proposed DSF-WLM algorithm with the complex angular central Gaussian mixture model (ACGMM) \cite{ito2016complex} and the original DSF by power mean (DSF-PM) on speech mixture recorded in three real acoustic environments. As with DSF-WLM, both ACGMM and DSF-PM are also initialized with KHL. Source extraction for all methods are performed using time-frequency masking. For ACGMM, we adopted the same mask defined in \cite{ito2016complex}. For DSF-based methods, we used the softargmax-based mask \cite{nguyen2020directional} defined by
\begin{equation}
    \mathcal{M}_{n,k} = \dfrac{\exp\left({
                -\beta \absv{\mbf{h}_n^\H \mbf{x}[k]}^2
            }/{
                \left(\norm{\mbf{h}_n} \norm{\mbf{x}[k]}\right)^2
            }\right)}{\sum_{i=1}^{N}\exp\left({
                -\beta \absv{\mbf{h}_i^\H \mbf{x}[k]}^2
            }/{
                \left(\norm{\mbf{h}_i} \norm{\mbf{x}[k]}\right)^2
            }\right)}, \label{eq:mask}
\end{equation}
where $\beta = 12.5$ is a softness parameter. Permutation alignment is performed using a modified version of the algorithm presented in \cite{wang2014multi} for all methods. We remark that the mask in \eqref{eq:mask} can be slightly suboptimal since it does not take the imbalance into account.

We recorded speech signals in three acoustic environments (an office, a lecture theatre, and a tutorial room) with different reverberation times. Within each acoustic environment, we recorded twelve speech source signals, each at a different location, as shown in Fig.~\ref{fig:setup}, using \SI{30}{\second} of concatenated utterances from twelve unique speakers in the TIMIT corpus \cite{garofolo1993darpa}. The source signals were recorded using a mobile phone with two built-in microphones spaced approximately \SI{24}{\milli\meter} apart at a sampling rate of \SI{16}{\kilo\hertz}. For each acoustic environment, we simulated mixtures of four sources, created by choosing a combination of four source locations, artificially introducing random attenuation of up to \SI{12}{\decibel} to all but one sources, and subsequently summing the four scaled source images to form a mixture. We used only fifty unique under-determined scenarios for evaluation per environment (out of 495 possible unique scenarios).

%

Based on our pilot study in a hold-out dataset collected from a different environment where the reverberation time $T_{60}$ estimated by \cite{lollmann2010improved} is approximately \SI{388}{\milli\second}, we found that DSF-WLM using $r$ that is close to $0$ achieves higher median performance compared to when $r$ is close to $1$. However, the latter provides more consistent separation performance. Fig.~\ref{fig:grid} shows the variation of SDR with $r$ for DSF-WLM with $\alpha=10$, which is also found via a grid search on the same dataset. For ease of comparison, we fixed the hyper-parameters $r = 0.5$, $\alpha = 10$ for DSF-WLM, and $p = -0.2$ for DSF-PM as these values achieved the highest performance on our hold-out dataset. All methods employ a \num{2048}-sample periodic Hamming window with \SI{75}{\percent} overlap for the short-time Fourier transform (STFT). In practice, these hyper-parameters, including those of STFT, should be optimized for each acoustic environment to achieve optimal performance.

Fig.~\ref{fig:results} shows the separation performance quantified via the source-to-distortion ratio (SDR) evaluated using the \texttt{BSS Eval v3.0} toolbox \cite{vincent2006performance} and the extended short-time objective intelligibility (ESTOI) measure \cite{jensen2016algorithm}. The reverberation times of the acoustic environments, estimated using \cite{lollmann2010improved}, have also been included in the figure. It can be seen that the proposed DSF-WLM algorithm generally exhibits improvements in SDR and ESTOI compared to the baselines. On the other hand, ACGMM consistently suffers from the worst performance in both SDR and ESTOI even though each source is given a weight parameter in the mixture model. As expected, the performance of all algorithms decreases with increasing $T_{60}$. However, it can be seen that DSF-based methods suffer less performance degradation compared to ACGMM as the $T_{60}$ increases. We note that DSF-WLM shows modestly lower median performance compared to DSF-PM in the tutorial room environment. This is due to a sub-optimal value of $r$ since the reverberation time of the tutorial room is nearly twice that of the hold-out environment where the hyper-parameters were optimized in.

It should also be noted that the proposed algorithm can be optimized using any unconstrained gradient methods whereas ACGMM relies on an expectation-maximization update rule which incurs significantly higher computational cost. Given similar stopping conditions, DSF-PM and DSF-WLM require approximately \SI{50}{\second} on average for a \num{30}-second mixture whereas ACGMM require approximately \SI{3}{\minute} on a machine with Intel Xeon Silver 4208 (\SI{2.10}{\giga\hertz}) CPU.

\section{CONCLUSION}
\label{sec:foot}

We proposed an algorithm based on the directional sparse filtering framework that is robust to imbalances in the source spectra by introducing the weighted Lehmer mean. Exploiting the simple gradient and weight self-normalization property, the Lehmer mean with learnable weights allow the proposed algorithm to adaptively account for source imbalances with minimal increase in computational complexity. Performance evaluation in multiple real acoustic environments show modest improvement in performance compared to the original DSF algorithm with power mean and considerable improvement compared to the complex angular central Gaussian mixture model.
\FloatBarrier
\bibliographystyle{IEEEbib}
\bibliography{IEEEabrv, refs}

\end{document}